# On Unconventional Electron Pairing In a Periodic Potential

Valentin Voroshilov, Physics Department, Boston University, Boston, MA, 02215, USA, valbu@bu.edu

On the assumption that two electrons with the same group velocity effectively attract each other a simple model Hamiltonian is proposed to question the existence of unconventional electron pairs formed by electrons in a strong periodic potential.

71.10.Li

Despite intensive experimental and theoretical research the mechanism responsible for high temperature superconductivity[1] is yet not clear. The goal of this letter is to offer some speculations on the matter as a step toward the future deeper theoretical analysis.

Among different theoretical approaches there is a view that formation of electron pairs is a natural property of interacting electrons immersed in a strong periodic potential[2,3]. In particular, as one of the possibilities a hole-electron paring mechanism is suggested, where an electron is paired with a hole traveling in space in opposite direction[4]. As a possible extension of that hole-electron paring mechanism we assume that electron pairs are formed by electrons *traveling with the same group velocity* (at this point we do not impose limits on the energy of interacting electrons).

Based on the assumption above, we write a model Hamiltonian in the following form (Eq. 1, 2):

$$H = \sum_{p\sigma}(\varepsilon_p - \mu)a^+_{p\sigma}a_{p\sigma} + U, \qquad \varepsilon_p = -t(\cos p_x + \cos p_y). \qquad (1)$$

In Eq. (1) the first term represents kinetic energy of electrons in a 2D square lattice in tight binding approximation; t > 0 is a hopping integral. Here and below we assume the lattice constant, the Plank's constant, and the Boltzmann's constant are equal to unity. For all electrons momenta lie in the first Brillouin zone $-\pi < p_{x,y} < \pi$; $\mu$ is a chemical potential introduced in the Hamiltonian as a Lagrange multiplier; $a^+_{p\sigma}$ and $a_{p\sigma}$ are creation and annihilation operators for an electron with momentum $p = (p_x, p_y)$ and spin component $\sigma$ ($\sigma = +, -$). Term $U$ in Eq. (1) describes an effective attraction between electrons with the same group velocity.

For an electron with momentum $p = (p_x, p_y)$ and kinetic energy $\varepsilon_p = -t(\cos p_x + \cos p_y)$ there might exists three more electrons with the same group velocity $v_p = \nabla \varepsilon_p$, namely,

$$p' = (p_x, \pi \cdot sgn(p_y) - p_y), \quad p'' = (\pi \cdot sgn(p_x) - p_x, p_y), \quad p^* = (\pi \cdot sgn(p_x) - p_x, \pi \cdot sgn(p_y) - p_y)$$

($sgn(p_{x,y})$ is the signum function).

Thus, in general, there might be formed a cluster of four electrons with the same group velocity. In this letter we limit ourselves to the *simplest* interaction term (Eq. 2)

$$U = -\frac{W}{N} V^+ V, \qquad V^+ = \sum_p a^+_{p\uparrow} a^+_{p^*\downarrow}. \tag{2}$$

In Eq. (2) $N$ is the number of sites, $-W < 0$ is a coupling constant for paired electrons, and the summation runs over the first Brillouin zone.

Let us introduce a Bogolyubov canonical transformation (Eq. 3)

$$a_{p\uparrow} = u_p b_{p\uparrow} - v_p b^+_{p^*\downarrow}, \quad a_{p\downarrow} = u_p b_{p\downarrow} + v_p b^+_{p^*\uparrow}, \quad u_p = u_{p^*} > 0, \quad v_p = v_{p^*} > 0, \quad u_p^2 + v_p^2 = 1. \tag{3}$$

With the means of transformation (3) the standard procedure[5] gives (Eq. 4) for the energy of the system as a function of the occupation numbers

$$H|E> = E|E>, \qquad b^+_{p\sigma} b_{p\sigma} |E> = n_{p\sigma} |E>, \quad n_{p\sigma} = 0,1, \tag{4a}$$

$$E = -2\mu \sum_p v_p^2 + \sum_{p,\sigma} (\varepsilon_p - \mu(u_p^2 - v_p^2)) n_{p\sigma} - \frac{W}{N} \left[ \sum_p u_p v_p (1 - n_{p+} - n_{p-}) \right]^2. \tag{4b}$$

Equation (5)

$$-2\mu u_p v_p = (u_p^2 - v_p^2) \frac{W}{N} \sum_p u_p v_p (1 - <n_{p+}> - <n_{p-}>) \tag{5}$$

insures minimization of the free energy of the system. This equation defines parameters $u_p$ and $v_p$ of canonical transformation (3).

Chemical potential can be related to the number of electrons as $N_e = -\frac{\partial <E>}{\partial \mu}$ which gives (Eq. 6)

$$N_e = 2\sum_p v_p^2 + \sum_p (u_p^2 - v_p^2)(<n_{p+}> + <n_{p-}>). \tag{6}$$

After elementary transformations we obtain (Eq. 7)

$$u_p = u = \frac{1}{\sqrt{2}}\sqrt{1-\frac{2\mu}{\Delta}}, \qquad v_p = v = \frac{1}{\sqrt{2}}\sqrt{1+\frac{2\mu}{\Delta}}, \qquad \Delta = \frac{W}{N}\sum_p(1-<n_{p+}>-<n_{p-}>) \qquad (7)$$

Solution (7) exist only when $\frac{2|\mu|}{|\Delta|} \leq 1$ (this is the only regime to be discussed below).

Combining Eq. (6) and (7) we find the chemical potential (Eq. 8)

$$\mu = \frac{N_e - N}{N} * \frac{W}{2}. \qquad (8)$$

For elementary excitations we obtain (Eq. 9)

$$n_p = <b^+_{p\sigma}b_{p\sigma}> = [1+\exp(\frac{\omega_p}{T})]^{-1}, \qquad (9a)$$

$$\omega_p = \varepsilon_p - \mu(u^2-v^2) + 2\frac{W}{N}u^2v^2\sum_p(1-<n_{p+}>-<n_{p-}>) = \varepsilon_p + \frac{\Delta}{2}. \qquad (9)$$

When for all momenta $\varepsilon_p + \frac{W}{2} > 0$ (i.e. when $\frac{W}{4t} > 1$, the only case we consider here) there are no excitations in the ground state ($n_p \equiv 0$ at $T = 0$); expression (9) shows the existence of a gap in the excitation spectrum, and pairing function $<a^+_{p+}a^+_{-p*+}> = -uv \neq 0$ shows the existence of electron pairs (even at a half filling when $N_e = N$, and $\mu = 0$). From the point of view of the BCS theory[6], one would conclude on the existence of superconductivity. However, at $T = 0$ electron density $<a^+_{p\sigma}a_{p\sigma}> = \frac{N_e}{2N}$ is a constant, which would mean that the Fermi see fills up the whole 1st Brillouin zone for all numbers of electrons. Correcting interaction term (2) (for example, by limiting attractive interactions being only between electrons with energy above the Fermi level), or/and including in the analysis four-electron clusters might lead to a physically more adequate picture.

There is also an interesting outlook, related to the notion that HTSC are doped Mott insulators[7]. The linkage between a parent insulating phase and a superconductive one is yet elusive and leaves a room for speculations.

The most general form of a Hamiltonian for a system of interacting non-relativistic electrons in an external field is given by a standard first quantization expression (Eq. 10)

$$H = \sum_{\alpha=1}^{N_e} \frac{\vec{p}_\alpha^2}{2m} + \sum_{\alpha=1}^{N_e} U(\vec{r}_\alpha) + \frac{1}{2}\sum_{\alpha=1}^{N_e}\sum_{\substack{\beta=1\\\beta\neq\alpha}}^{N_e} V(\vec{r}_\alpha,\vec{r}_\beta), \qquad V(\vec{r}_\alpha,\vec{r}_\beta) = V(\vec{r}_\beta,\vec{r}_\alpha). \qquad (10)$$

This expression can be rewritten in a different form (Eq. 11)

$$H = H_0 + H_{int}, \qquad H_0 = \sum_{\alpha=1}^{N_e} \frac{\vec{p}_\alpha^2}{2m} + \sum_{\alpha=1}^{N_e} U_{eff}(\vec{r}_\alpha), \qquad H_{int} = \frac{1}{2}\sum_{\alpha=1}^{N_e}\sum_{\substack{\beta=1\\\beta\neq\alpha}}^{N_e} V_{eff}(\vec{r}_\alpha,\vec{r}_\beta), \qquad (11a)$$

$$U_{eff}(\vec{r}_\alpha) = U(\vec{r}_\alpha) + W(\vec{r}_\alpha), \qquad V_{eff}(\vec{r}_\alpha,\vec{r}_\beta) = V(\vec{r}_\alpha,\vec{r}_\beta) - \frac{W(\vec{r}_\alpha) + W(\vec{r}_\beta)}{N_e - 1}. \qquad (11b)$$

For any function $W(\vec{r}_\alpha)$ Hamiltonian (11) is exactly equal to Hamiltonian (10). However, if one intends on using a perturbation theory, the difference may arise from different sets of eigenfunctions for one-particle operators using to write the Hamiltonian in a second quantization form (and the use of $V_{eff}(\vec{r}_\alpha,\vec{r}_\beta)$ can provide a better convergence of the perturbation series).

In a mean-field approach function $W(\vec{r}_\alpha)$ can be seen as an effective energy of a single electron in the filed of all other electrons.

When electrons are localized because of strong interactions, the inclusion of function $W(\vec{r}_\alpha)$ might make a significant difference. If function $W(\vec{r}_\alpha)$ has the same symmetry as lattice potential $U(\vec{r}_\alpha)$ has, the effective potential $U_{eff}(\vec{r}_\alpha)$ should lead to the band structure similar to the one provided by a "naked" lattice potential $U(\vec{r}_\alpha)$. Although, the parameters of the system can be modulated, for example, "potential" $W(\vec{r}_\alpha)$ can increase the potential barriers of potential $U(\vec{r}_\alpha)$, which effectively changes hopping constant $t$ (i.e. the width of the band).

If the symmetry of term $W(\vec{r}_\alpha)$ is different from the symmetry of lattice potential $U(\vec{r}_\alpha)$ (i.e. the spatial symmetry of the electron system is different from the spatial symmetry of the lattice), that difference might lead to important changes in the structure of energy bands.

In particular, if the effective field of electrons, which is also strong, is of a different symmetry than the lattice field, in result, even if formally the lattice is at a half filling, effectively electrons can occupy "pseudo sites" by two on each such a site (a "pseudo lattice"). Such a dynamic symmetry braking may lead to an insulating phase where the band theory predicts a metallic one.

Under an assumption that copper-oxides might experience such a symmetry breaking, one could argue that in model (1) at a formal half-filling when $N_e = N$, the effective number of sites $N_{eff}$ is to be equal to a half the number of actual sites $N$; and, hence, in all expressions (2) – (9) $N$ should be replaced with $N_{eff} = \frac{N_e}{2}$. This leads (at $T = 0$) to $\mu = \frac{\Delta}{2} = \frac{W}{2}$, the energy gap still exists, but the pairing function $< a^+_{p+} a^+_{p^*+} > = 0$ (since $u = 0$). This might be seen as a sign for the existence of a pseudo-gap.

In the end we briefly examine the existence of pairs when attractive interactions are limited between only some of the electrons.

Let us assume that at $T = 0$ electron distribution in momentum space is of simplest form (Eq. 12)

$$< a^+_{p\sigma} a_{p\sigma} > = \begin{cases} 1, & \varepsilon_p < \varepsilon_F \\ 0, & \varepsilon_p > \varepsilon_F \end{cases}, \qquad -2t < \varepsilon_F < 2t. \qquad (12)$$

We also assume that in term (2) two electrons experience attraction only when they both have energy above Fermi level $\varepsilon_F$. Under these assumptions Fermi level has to be negative $\varepsilon_F < 0$; and we have to modify the Hamiltonian and write it in the following form (Eq. 13)

$$H = \sum_{\substack{p\sigma \\ \varepsilon_p < \varepsilon_F \\ -\varepsilon_F < \varepsilon_p}} (\varepsilon_p - \mu) a^+_{p\sigma} a_{p\sigma} + \sum_{\substack{p\sigma \\ \varepsilon_F < \varepsilon_p < -\varepsilon_F}} (\varepsilon_p - \mu) a^+_{p\sigma} a_{p\sigma} - \frac{W}{N} V^+ V, \qquad V^+ = \sum_{\substack{p \\ \varepsilon_F < \varepsilon_p < -\varepsilon_F}} a^+_{p\uparrow} a^+_{p^*\downarrow}. \qquad (13)$$

Canonical transformation (3) remains applied to the operators with momenta within region $\varepsilon_F < \varepsilon_p < -\varepsilon_F$, and outside this region we can set $a_{p\sigma} = b_{p\sigma}$.

Calculations show that solution (7) remains correct for $u$ and $v$, but the summation in the expression for $\Delta$ should be limited by momenta for which $\varepsilon_F < \varepsilon_p < -\varepsilon_F$. Expression (8) for the chemical potential stays but now it is also equal to the Fermi level, so $\varepsilon_F = \mu = \frac{N_e - N}{N} * \frac{W}{2}$; which means $N_e < N$, and $\frac{N - N_e}{N} * \frac{W}{4t} < 1$.

Expression (9) is to be replaced by

$$n_p = <b^+_{p\sigma} b_{p\sigma}> = \begin{cases} [1+\exp(\frac{\varepsilon_p - \mu}{T})]^{-1}, & \varepsilon_p < \varepsilon_F, \quad -\varepsilon_F < \varepsilon_p \\ [1+\exp(\frac{\varepsilon_p + \frac{\Delta}{2}}{T})]^{-1}, & \varepsilon_p < \varepsilon_p < -\varepsilon_F \end{cases}. \qquad (14)$$

Since by the assumptions there are no excitations in the ground state, hence $-2t + \frac{\Delta_{(T=0)}}{2} > 0$, which provides an additional condition on the parameters of the system; $\frac{W}{4tN} \sum_{\substack{p \\ \varepsilon_F < \varepsilon_p < -\varepsilon_F}} 1 > 1$.

When $-\mu < \frac{\Delta_{(T=0)}}{2}$ the excitation spectrum has an energy gap. What differs this model from a BCS picture is that parameter $\Delta$ is never equal to zero (for all the values of the parameters when solution (14) exists and consistent with condition (12)). However, when the temperature of the system is such that $\Delta = \frac{-\mu}{2}$ solution (7) gives $u = 1$, $v = 0$, and pairing function becomes zero $<a^+_{p+} a^+_{p^*+}> = 0$. What kind of a transition is happening at that temperature is a matter of a subsequent study.

We see that a model based on Hamiltonian (1), (2), or (13) shows unusual and interesting properties. At this point there is no solid theoretical basis to support the speculations on relationships between the model and properties of HTSC. The future work will show if such a basis exists.

**Appendix**

Let us write Schrödinger equation for two non-relativistic electrons with anti-parallel spins (all the constants a set to unity, terms are rearranged)

$$\left( \frac{\partial^2}{\partial \vec{r}_1^2} + \frac{\partial^2}{\partial \vec{r}_2^2} + 2(E - V(\vec{r}_1) - V(\vec{r}_2)) \right) \Psi(\vec{r}_1, \vec{r}_2) = 2U(\vec{r}_1, \vec{r}_2) \Psi(\vec{r}_1, \vec{r}_2),$$

where $U(\vec{r}_2, \vec{r}_2)$ describes the Coulomb's interaction and $V(\vec{r})$ describes an external field.

The standard approach is to seek the solution with the means of a Green's function

$$\Psi(\vec{r}_1, \vec{r}_2) = \Phi(\vec{r}_1, \vec{r}_2) + 2 \iint d\vec{r}_1^{'} d\vec{r}_2^{'} G(\vec{r}_1, \vec{r}_2; \vec{r}_1^{'}, \vec{r}_2^{'}) U(\vec{r}_1^{'}, \vec{r}_2^{'}) \Psi(\vec{r}_1^{'}, \vec{r}_2^{'})$$

where "free" solution $\Phi(\vec{r}_1,\vec{r}_2)$ and Green's function $G(\vec{r}_1,\vec{r}_2;\vec{r}_1',\vec{r}_2')$ satisfy equations

$$\left(\frac{\partial^2}{\partial \vec{r}_1^2} + \frac{\partial^2}{\partial \vec{r}_2^2} + 2(E - V(\vec{r}_1) - V(\vec{r}_2))\right)\Phi(\vec{r}_1,\vec{r}_2) = 0$$

$$\left(\frac{\partial^2}{\partial \vec{r}_1^2} + \frac{\partial^2}{\partial \vec{r}_2^2} + 2(E - V(\vec{r}_1) - V(\vec{r}_2))\right)G(\vec{r}_1,\vec{r}_2;\vec{r}_1',\vec{r}_2') = \delta(\vec{r}_1 - \vec{r}_1')\delta(\vec{r}_2 - \vec{r}_2').$$

"Free" solution $\Phi(\vec{r}_1,\vec{r}_2)$ can be represented as

$$\Phi(\vec{r}_1,\vec{r}_2) = \varphi_{\vec{p}}(\vec{r}_1)\varphi_{\vec{q}}(\vec{r}_2) + \varphi_{\vec{q}}(\vec{r}_1)\varphi_{\vec{p}}(\vec{r}_2)$$

where

$$\left(\frac{\partial^2}{\partial \vec{r}} + 2(E_{\vec{p}} - V(\vec{r}))\right)\varphi_{\vec{p}}(\vec{r}) = 0 \quad \text{and} \quad E = E_{\vec{p}} + E_{\vec{q}}.$$

Functions $\varphi_{\vec{p}}(\vec{r})$ can be normalized as $\sum_{\vec{p}} \varphi_{\vec{p}}^*(\vec{r})\varphi_{\vec{p}}(\vec{r}') = \delta(\vec{r} - \vec{r}')$.

In that case the Green's function can be written as

$$G(\vec{r}_1,\vec{r}_2;\vec{r}_1',\vec{r}_2') = \frac{1}{2}\sum_{\vec{p}_1,\vec{p}_2} \frac{\varphi_{\vec{p}_1}(\vec{r}_1)\varphi_{\vec{p}_1}^*(\vec{r}_1')\varphi_{\vec{p}_2}(\vec{r}_2)\varphi_{\vec{p}_2}^*(\vec{r}_2')}{E_{\vec{p}} + E_{\vec{q}} - E_{\vec{p}_1} - E_{\vec{p}_2}}.$$

For a strong periodic external field the approximate solution (for example tight binding approximation) gives $E_{\vec{p}} \approx \alpha - t(\cos(p_x) + \cos(p_y)) = \alpha + \varepsilon_{\vec{p}}$ (we consider only 2D square lattice). In that case the Green's function can be rewritten as

$$G(\vec{r}_1,\vec{r}_2;\vec{r}_1',\vec{r}_2') = \frac{1}{2}\sum_{\vec{p}_1,\vec{p}_2} \frac{\varphi_{\vec{p}_1}(\vec{r}_1)\varphi_{\vec{p}_1}^*(\vec{r}_1')\varphi_{\vec{p}_2}(\vec{r}_2)\varphi_{\vec{p}_2}^*(\vec{r}_2')}{\varepsilon_{\vec{p}} + \varepsilon_{\vec{q}} - \varepsilon_{\vec{p}_1} - \varepsilon_{\vec{p}_2}}.$$

It is clear that for this solution the Green's function exhibits irregular behavior (which is unique for "energy spectrum" $\varepsilon_{\vec{p}} = -t(\cos(p_x) + \cos(p_y))$), namely, there are momenta $\vec{p}$ and $\vec{q}$ for which $\varepsilon_{\vec{p}} + \varepsilon_{\vec{q}} = 0$. It happens, in particular, when $\vec{q} = \vec{p}^* = (\pi \cdot sgn(p_x) - p_x, \pi \cdot sgn(p_y) - p_y)$, i.e. when two "free" electrons have the same group velocity. Further investigation is needed to examine connection between this irregularity and electron pairing in HTSC.

It might be plausible to offer a "naïve" picture for two electrons paired in a strong periodic potential. When two electrons are localized at neighboring sites (having opposite spins), their exchange interaction (which takes much longer time than for almost free electrons) results in strong correlations between the electrons, and the paired electrons have a tendency traveling together by hopping simultaneously unto neighboring sites in the same direction. This correlation might make electrons with the same group velocity being special, and might have a significant influence on properties of the system at the optimal electron density (i.e. enough electrons to pair, and enough empty sites for hopping).